\documentclass[epsfig,showpacs,floatfix]{revtex4}
\usepackage{epsfig,color}
\usepackage{epsfig}
\usepackage{lscape}
\usepackage{amsmath}
\usepackage{amssymb}
\usepackage{graphicx}
\usepackage{graphics}
\usepackage{subfigure}

\def\bea{\begin{eqnarray}}
\def\eea{\end{eqnarray}}
\def\be{\begin{equation}}
\def\ee{\end{equation}}

\begin{document}

\title{Identified Particle Spectra  for Au+Au Collisions at $\sqrt{s}$ = 200 GeV from \\
        STAR, PHENIX and BRAHMS in Comparison to Core-Corona Model Predictions}

\author{C. Schreiber, K. Werner and J. Aichelin\footnote{invited speaker}}

\address{SUBATECH, Universit\'e de Nantes, EMN, IN2P3/CNRS
\\ 4 rue Alfred Kastler, 44307 Nantes cedex 3, France}
\begin{abstract}
The core-corona model describes quite successfully the centrality dependence of multiplicity and $<p_t>$ of identified particles observed in heavy ion reaction at beam energies between $\sqrt{s}$ = 17 GeV and 200 GeV. Also the centrality dependence of the elliptic flow, $v_2$, for all charged and identified particles could be explained in this model. Here we extend this analysis  and study the centrality dependence of single particle spectra of identified particles. We concentrate here on protons, antiprotons, kaons and pions which have all been measured by the PHENIX, STAR and BRAHMS collaborations. We find that an analysis of the spectra in the core-corona model suffers from differences in the data published by the different experimental groups, notably for the pp collisions. For each experience the data agree well with the prediction of the core-corona model but the value of the two necessary parameters depends onthe experiments.
\end{abstract}
\maketitle

\section{Motivations}
Lattice calculations predict that at high density and temperature hadronic matter transforms into a plasma of quarks and gluons. There is evidence that such a state can be obtained in heavy ions collisions at beam energies which can be reached at the colliders at CERN and in Brookhaven. This plasma is a very short living state - it lasts less than $10^{-23}$ seconds -  but it is assumed that this time is sufficiently long for reaching equilibrium. Hydrodynamical calculations indeed show that 
the experimental spectra can be described assuming that an equilibrium is obtained in a very short time. The origin of such a fast equilibrium is still discussed. The matter expands quickly toward the chiral/confinement phase transition during which
hadrons are formed. It came as a surprise that the multiplicity of identified stable particles in the most central collisions 
agrees almost perfectly with that expected for a statistical distribution at a freeze out temperature of around 170 MeV and a small baryon chemical potential. This has been taken as well as evidence that at least at the chiral/confinement phase transition the system is in equilibrium.

For symmetric systems the number of projectile participants equals that of target participants, independent of the centrality. If each particpant contributes the same energy in the center of mass system and if the system come to equilibrium one does not expect that the multiplicity per participating nucleon varies with centrality.  In the experiments such a variation has been observed, however.  In addition it depends strongly on the particle species. Whereas for $\pi$ this ratio is almost constant, for multi strange baryons this ratio varies by a large factor, a phenomenon which has been dubbed strangeness enhancement.

One of the crucial assumptions in statistical model calculations is that geometry does not play a role and {\it all} nucleons
come to statistical equilibrium. In simulations of the heavy ion reactions it has been observed that especially those nucleons
which are at the edge of the overlap region have initially few, in some cases only one, collision \cite{Werner:2007bf}. It does not
seem to be realistic that those equilibrate. The ratio of these edge nucleons as compared to all participating nucleons decrease with centrality. They do not present an important fraction for the most central collisions. This observation has motivated the core-corona model in which is it assumed that nucleons which scatter initially only once (corona particles)  are not part of the equilibrated source but produce particles as in pp collisions whereas all the other come to statistical equilibrium (core particles). Of course this fast transition between core and corona particles is a crude approximation but it allows to define from experimental pp and central AA data the centrality dependence of the different observables. Studies have shown that the present quality of data does not allow for a more refined definition of the transition between core and corona particles.
It has been further verified that the core-corona model describes quantitatively the results of the much more involved EPOS
simulation program. 

In a series of papers \cite{Aichelin:2008mi,Aichelin:2010ed,Aichelin:2010ns} it has been shown that the core-corona model describes quite nicely the centrality dependence of
the multiplicity of identified particles,  $<p_t>$ of identified particles and even of $v_2$ observed in AuAu and PbPb collisions.
The latter has been considered as a test ground for the shear viscosity needed to describe heavy ion data in viscous hydrodynamical calculations. The core-corona model describes this data without any reference to a viscosity.
The prediction for the CuCu data are completely determined by the AuAu data and agree with data as far as data have been published.

In this contribution we go one step further and study the observed single particle $p_t$ spectra for antiprotons, protons, kaons and pions. This study has the following objectives:\\
\noindent
a) To investigate whether the core-corona model, which has been successfully applied to described the centrality dependence of several averaged variables of identified and not identified particles like the multiplicity,  $<p_t>$ and the elliptic flow $v_2$ , is also able to describe the centrality dependence of the spectra\\
b) Experimentally is has been observed that the centrality dependence of $<p_t>$ for $\pi$, K, p and ${\bar p }$ is rather different. This has been interpreted as a sign for collective radial flow which yields a stronger increase of $<p_t>$ with centrality for heavy hadrons as compared to light ones. Here we address the question whether this increase is really only due to the superposition of core and corona spectrum or whether the increase of the mean values signals deviations from a simple superposition. 
\\
c)  To search for domains in $p_t$ which show deviations from the core-corona prediction, expected if interactions between core and corona particles become important,  and to try to interpret those deviations, if they exist, in physical terms.

\section{Centrality and core-corona fraction}
In order to determine the centrality dependence of the specta we have first of all to know the relative contribution of core and corona particles as a function of the centrality.The core-corona model relies on a single parameter : $f(N_{core})$, the fraction of core nucleons as a function of the centrality. Along with the number of participans, $N_{part}$, it is calculated by a Monte-Carlo simulation based on a Glauber model for hadrons in the nucleus. The parameters of the Glauber distribution are the only freedom of this model. We apply here the EPOS approach. The results are presented in table 1.
\begin{table}[h]
\begin{tabular}{ r | c | c | c | c | c | c | c | c | c | c }
Centrality & 80\%-92\% & 70\%-80\% & 60\%-70\% & 50\%-60\% & 40\%-50\% & 30\%-40\% & 20\%-30\% & 10\%-20\% & 5\%-10\% & 0\%-5\% \\ \hline
$N_{part}^{EPOS}$ & 7.939 & 16.40 & 29.87 & 51.01 & 80.32 & 119.0 & 169.1 & 235.1 & 300.9 & 351.3 \\ \hline
$f_{core}(N_{part})$ & 0.269 & 0.407 & 0.504 & 0.587 & 0.669 & 0.727 & 0.782 & 0.831 & 0.869 & 0.911 \\ \hline
$N_{part}^{STAR}$ &  & 10.7 & 22.0 & 40.6 & 67.8 & 105.4 & 155.9 & 223.6 & 289.6 & 345.8 \\ \hline
$\Delta N_{part}^{STAR}$ &  & 2.1 & 3.2 &4.3  & 5.0 & 5.3 & 5.1 & 4.2 & 3.0 & 1.9 \\ \hline
$N_{part}^{PHENIX}$           & 6.3 & 13.4 & 25.7 & 45.5 & 74.4 & 114.2 & 166.6 & 234.6 & 299.0 & 351.4 \\ \hline
$\Delta N_{part}^{PHENIX}$ &1.2  & 3.0  & 3.8 &3.3  & 3.8 & 4.4 & 5.4 & 4.7 & 3.8 & 2.9 \\ \hline
\end{tabular}
\label{tbl:fcore}
\caption{Centrality dependence of $f(N_{core})$ and $N_{part}$ for STAR, PHENIX and the core-corona model.}
\end{table}
We present there for the different centrality bins the average number of participants,  $N_{part }$, as well as the core fraction $f_{core}$. The STAR \cite{Abelev:2008zk} and PHENIX  \cite{Adler:2003cb} collaboration found other and mutually different values of $N_{part }$ for the same centrality bins. The values of the two collaborations agree within error bars but since this is a completely theoretical quantity  it is not clear why the difference cannot be avoided. Because different $N_{part}$ yield different $f_{core}$ it becomes difficult to compare the different experiments with the same model parameters.
\section{Results}
\subsection{Multiplicity}
\begin{figure}[h]
\centering
\subfigure[Core - Corona $N_{part}$]
{
\includegraphics[scale=0.5]{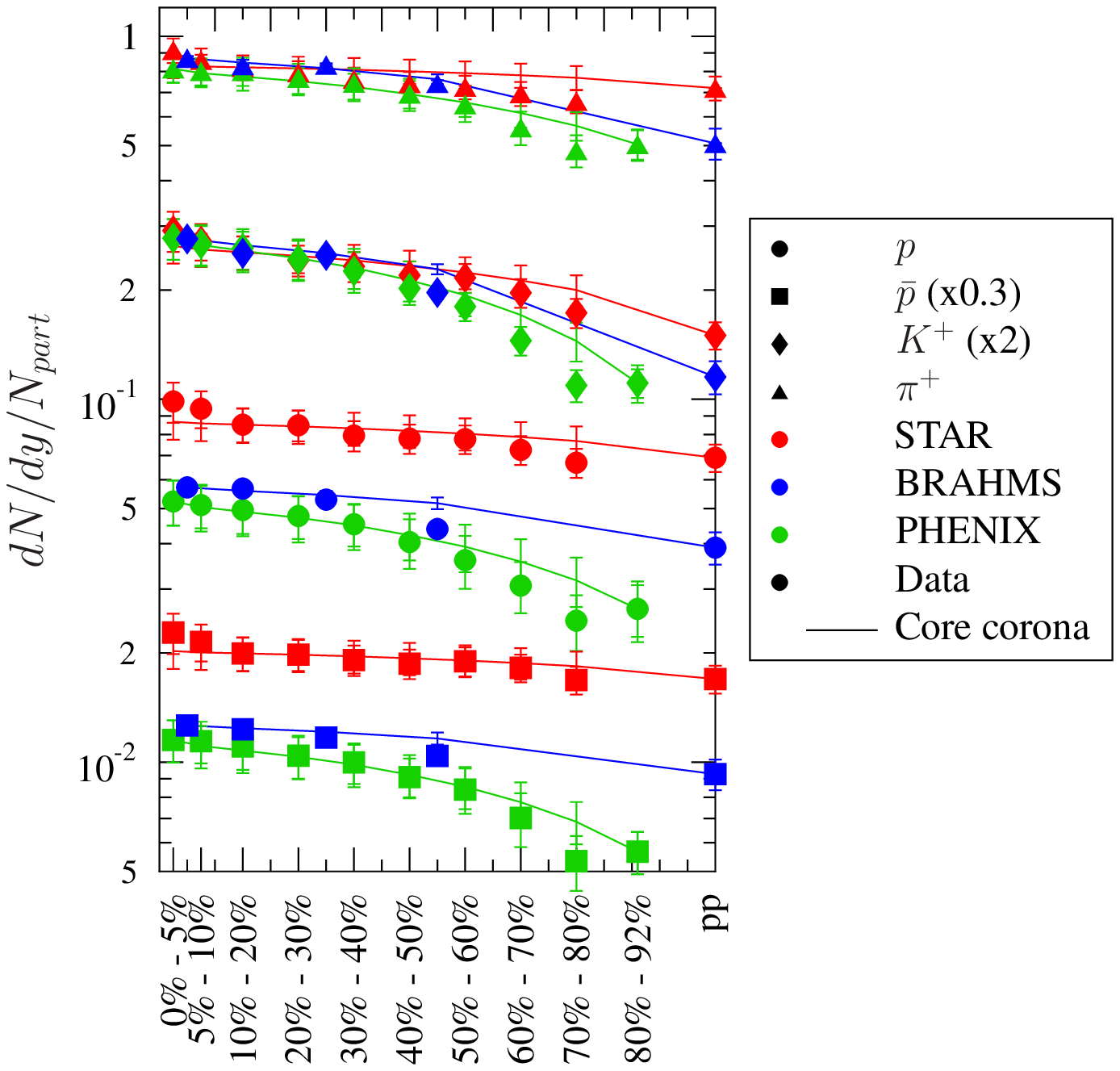}
\label{fig:multCC}
}
\subfigure[Respective experimental $N_{part}$]
{
\includegraphics[scale=0.5]{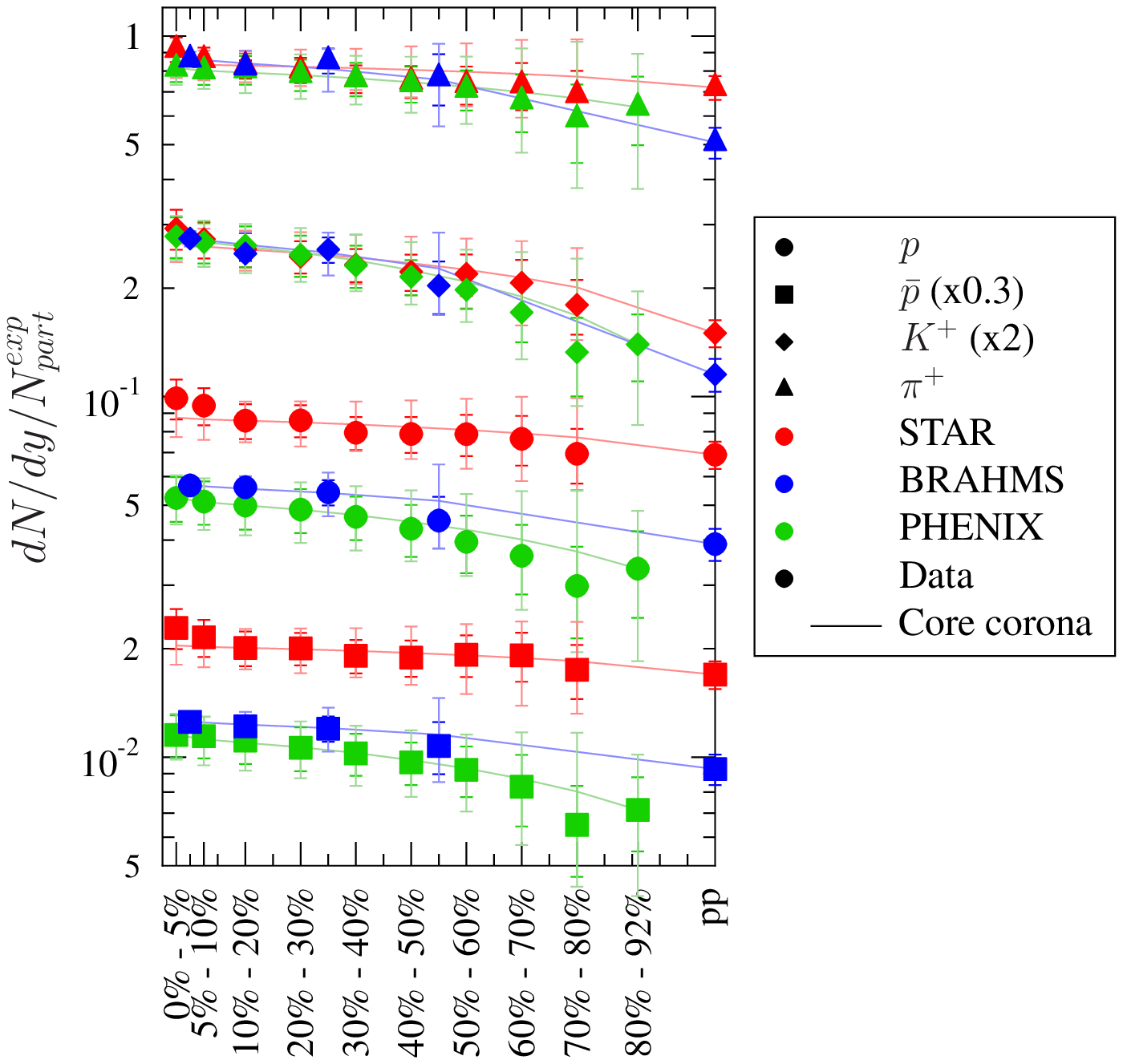}
\label{fig:multResp}
}
\caption{Multiplicity per participant as a function of centrality. On the left hand side we use $N_{part}$ as calculated by EPOS, on the right hand side the different $N_{part}$ quoted by the experiments for the given centrality bin. Symbols are data, the core-corona predictions are displayed as lines.}
\end{figure}
In this contribution we investigate the centrality dependence of the $\pi$, K and proton data measured by all three RHIC collaborations, STAR \cite{:2008ez}, PHENIX  \cite{Adler:2003cb} and Brahms \cite{Adler:2003cb,YangPhD} . According to the publications statistical errors are small and therefore systematical errors dominate. The spectra of $\pi$ and K of the three experiments should be directly comparable. For the proton spectra two collaborations (BRAHMS and PHENIX)  have corrected the spectra for feed down due to weak decay whereas STAR published uncorrected data. This is regrettable in view of the fact that the vertex resolution of STAR is better than that of PHENIX. Brahms, however, did not apply a feed down correction for the preliminary proton spectrum in pp collision yet \cite{YangPhD}. In order to apply the core-corona model to the Brahms data we have assumed that 40\% of the protons in pp are due to weak decay, employing the reduction factor published by the PHENIX collaboration \cite{Adler:2003cb}.   

Fig. 1 displays the experimental data. On the left hand side we have divided the published value of the multiplicity by $N_{part}$ calculated in the EPOS approach for the concerned centrality bin (see table 1). On the right hand side we have divided  the multiplicity by the $N_{part}$ value which has been published by the experimental groups.
We observe that for $\pi$ and K the multiplicities for central events agree quite well and are in the mutual error bars
but if one looks closer one realizes that STAR observes a increase of 10\% of the multiplicity of the pions toward the most central bin which is not seen by PHENIX. Toward peripheral collisions the differences increase and the data are outside the mutual error bars. For the multiplicities observed for pp resp. very peripheral AA ($N_{part}$=6.9) collisions we observe differences of 30\%. The STAR data set is almost compatible with a centrality independent $\pi$ multiplicity per participant, whereas for the PHENIX data set this is certainly not the case. If one takes the different $N_{part}$ values of the different experiments, right,  the differences become smaller. 

For $p$ and ${\bar p}$ the heavy ion data of BRAHMS and PHENIX agree but the preliminary pp data points of BRAHMS are almost a factor of 2 higher than that for the most peripheral data point of PHENIX.  In ref. \cite{Adler:2003cb} the PHENIX collaboration quotes that  between 30\% and 40\% of the observed protons and anti-protons are due to weak decay.  Therefore it is surprising that the multiplicity observed by  STAR and PHENIX differ by more than a factor of 2 in central collisions and by an even larger factor (close to 3) for peripheral reactions where the correction due to weak decay should be smaller (strangeness enhancement). Also the trends are quite different. Whereas STAR and BRAHMS observe a quite modest
centrality dependence PHENIX observes a quite strong one.

In the core-corona model the centrality dependence of the multiplicity  of a given particle species $i$ in a centrality bin containing $N_{part}$ participants is given by :
\begin{align}
M^i(N_{part}) &= N_{part}.\left[f_{core}(N_{part}).M^i_{core} + (1 - f_{core}(N_{part})).M^i_{corona}\right]
\label{eq:M}
\end{align}
where $M^i_{core}$ is the multiplicity per core participant and $M^i_{corona}$ the multiplicity per corona participant. There are several ways to determinate these two values: one can either calculate them from integrated fits or one can use directly the  published values, which are the results from a fit of a specific form (blast wave model) to the experimental spectra. We chose the latter in the present study: we use the multiplicity measured in pp and divided by a factor of two for $M^i_{corona}$. Then we extract $M^i_{core}$ from the most central multiplicity using eq. (1). If there was no suitable pp data like for  PHENIX, we use instead the most peripheral AA bin and eq. (3) to determine $M^i_{corona}$.
The result is displayed in fig.  1 as straight line. We observe that for all experiments the centrality dependence is that which is expected in the core-corona model but, as already said, the large difference between the experiments does not allow for  an unique value of  $M^i_{corona}$ and of $M^i_{core}$.

\subsection{Spectra}
\begin{figure}
\includegraphics[width=0.3\textwidth]{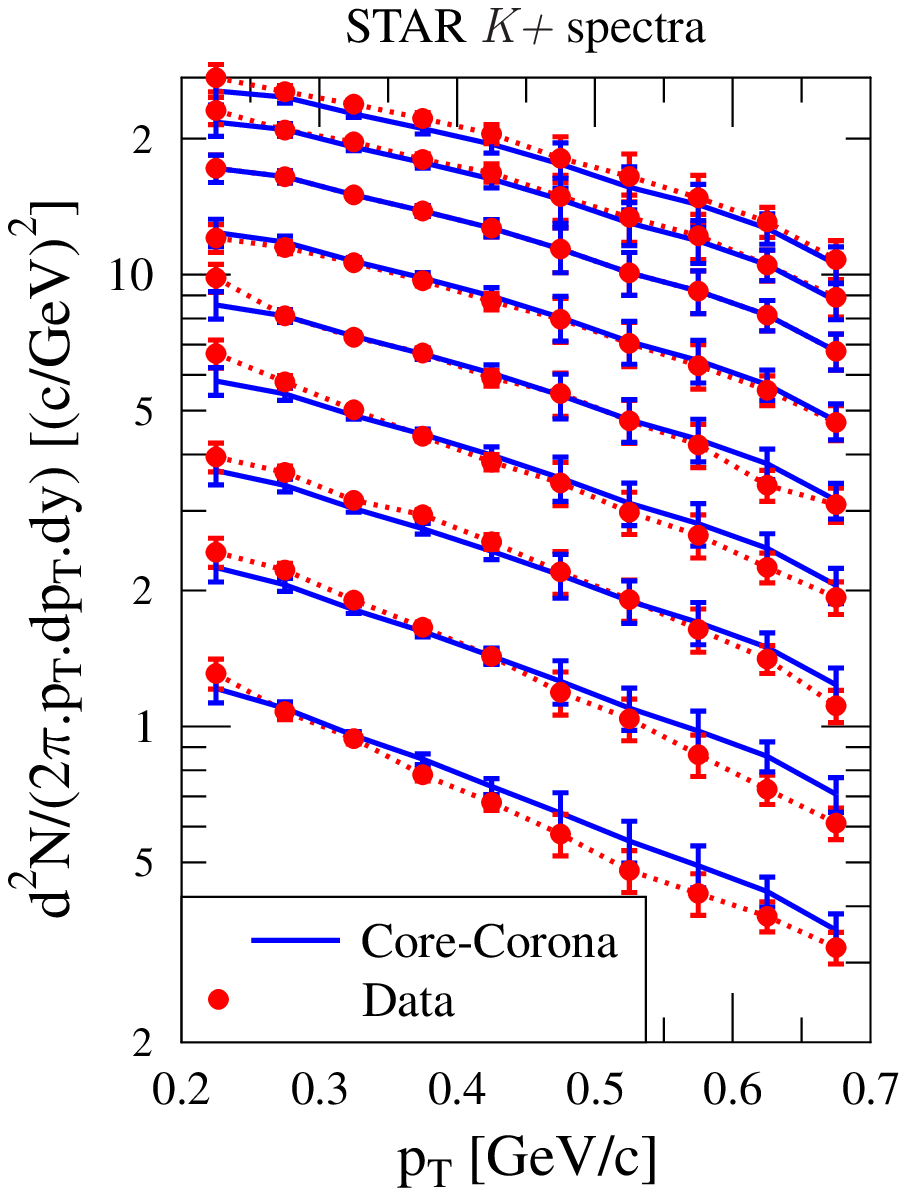}
\includegraphics[width=0.3\textwidth]{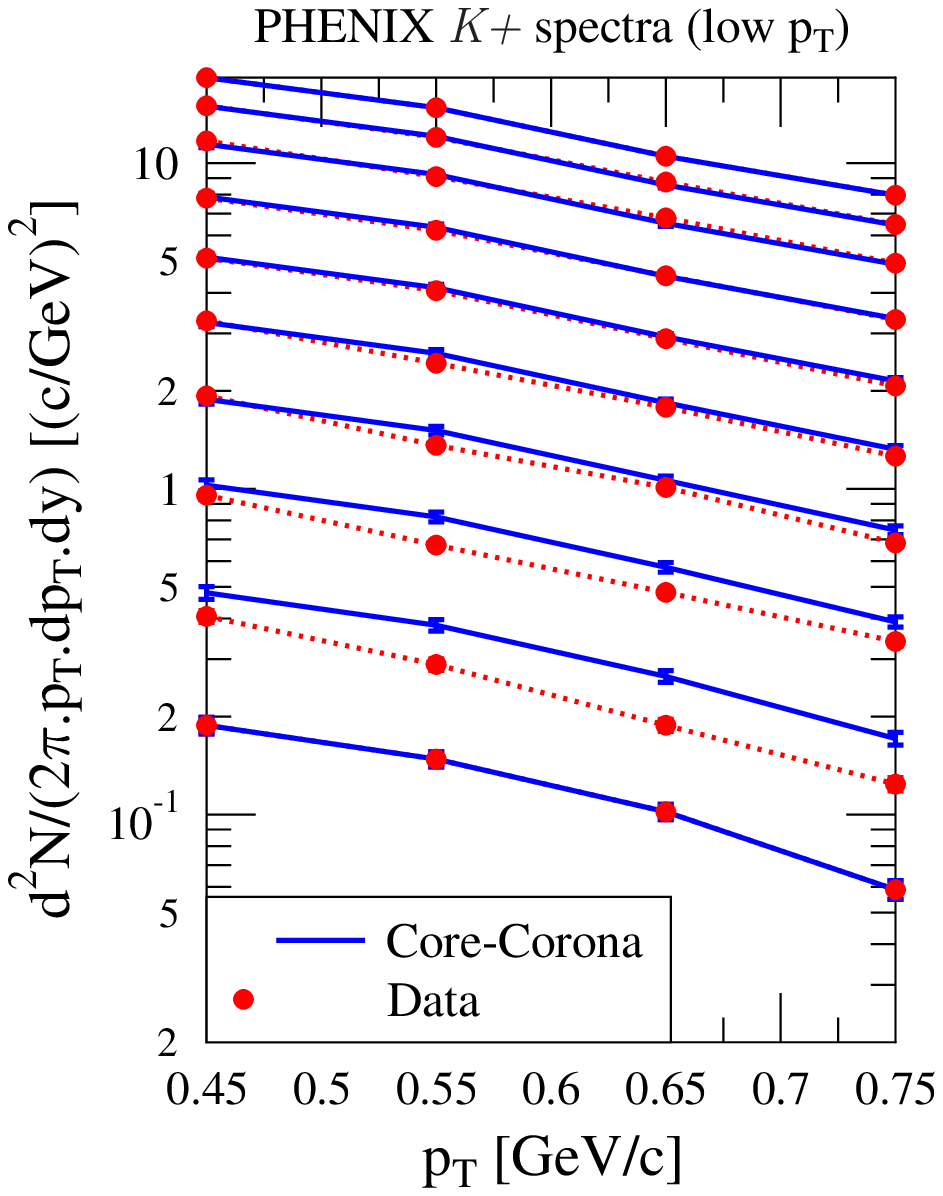}
\includegraphics[width=0.3\textwidth]{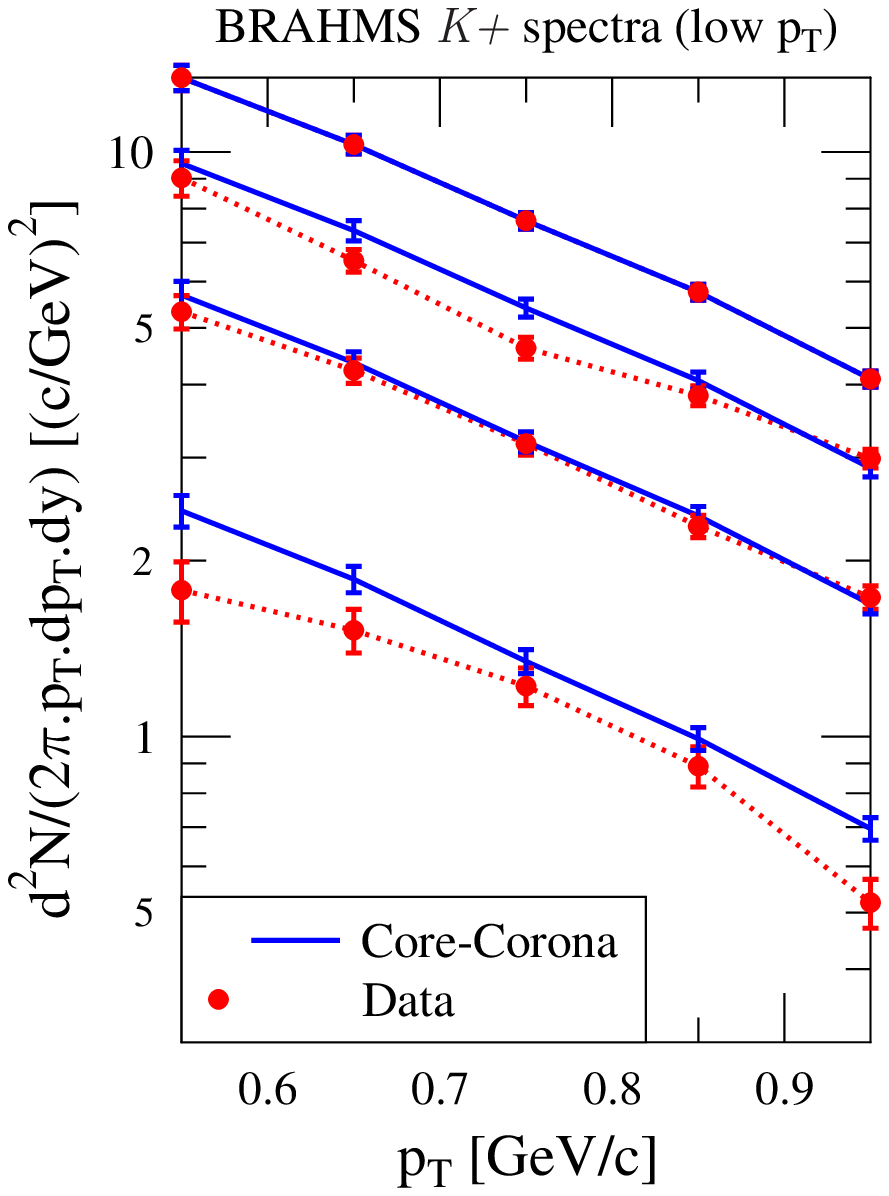}

\includegraphics[width=0.3\textwidth]{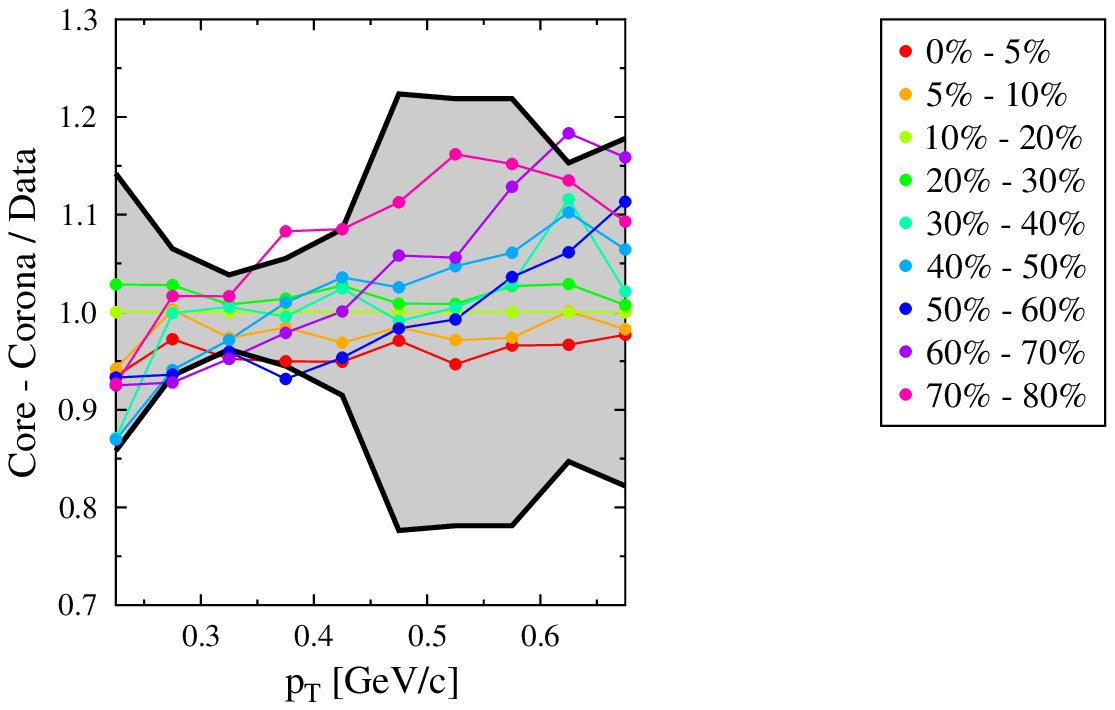}
\includegraphics[width=0.3\textwidth]{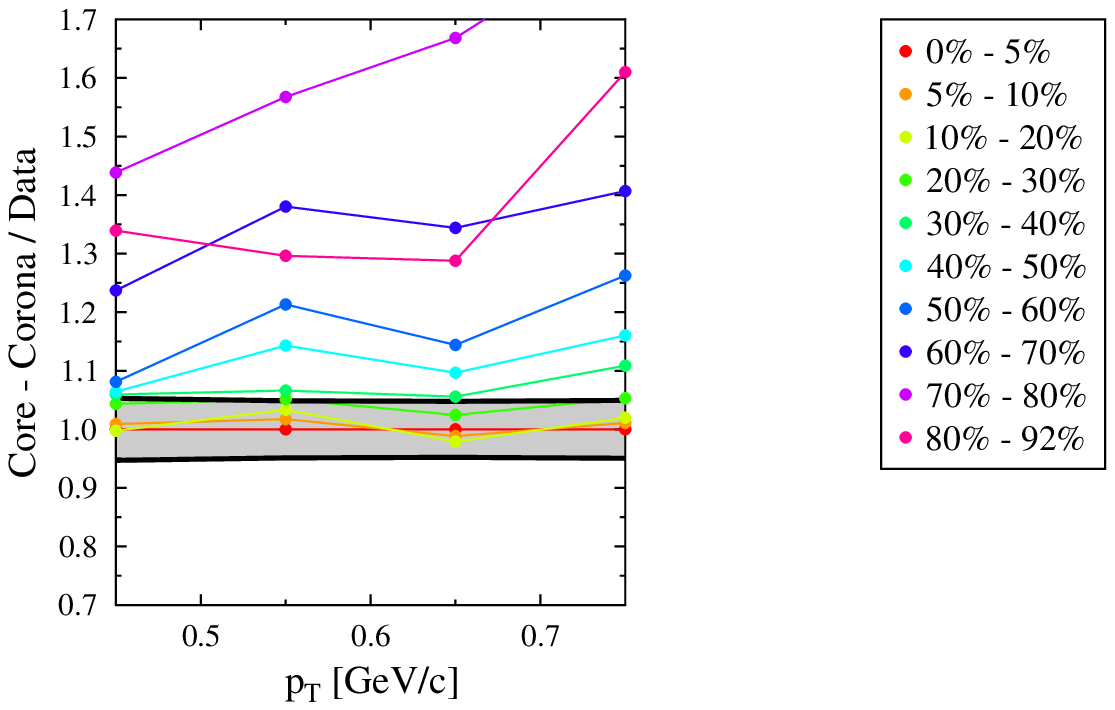}
\includegraphics[width=0.3\textwidth]{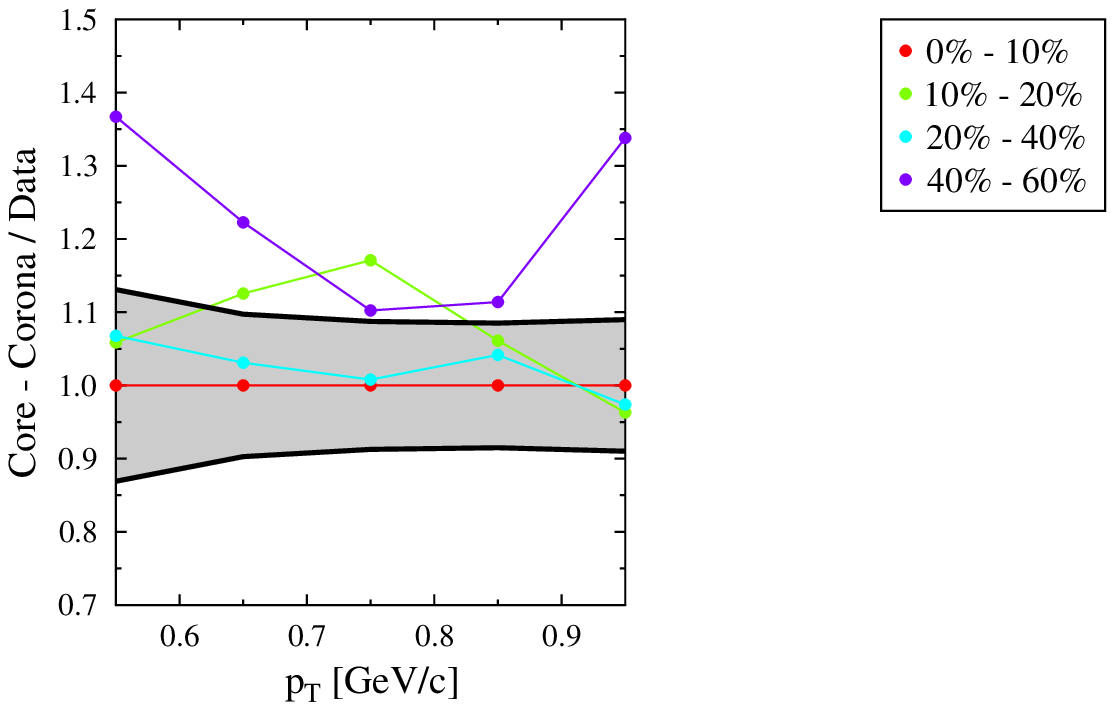}

\includegraphics[width=0.3\textwidth]{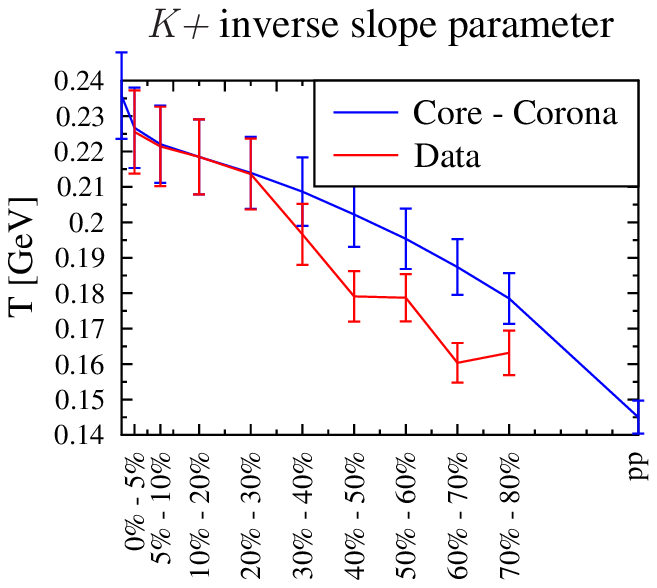}
\includegraphics[width=0.3\textwidth]{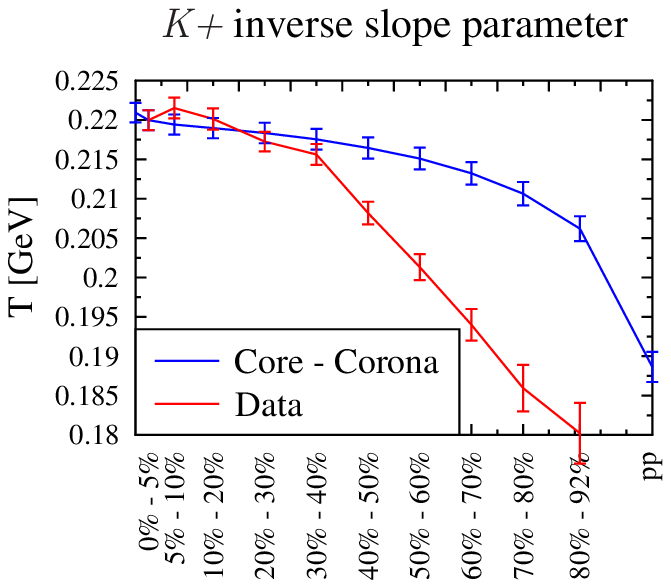}
\includegraphics[width=0.3\textwidth]{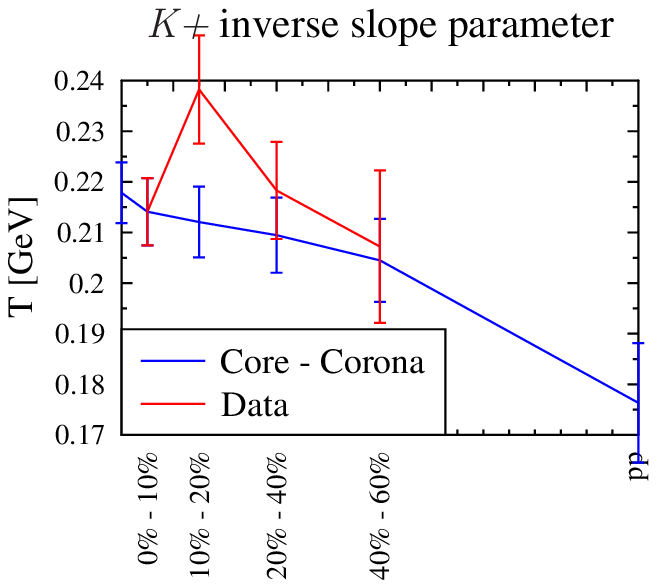}
\caption{Centrality dependence of the $K^+$ spectra measured by the STAR (left), PHENIX (middle) and BRAHMS (right) collaborations in comparison with the predictions of the core-corona model. Top:
The experimental spectra in comparison with the model prediciton. Middle: Ratio of the prediction of the
core-corona model and the experimental data. The shaded area mark the experimental errors. Bottom: Inverse slope parameter T obtained by fitting data and theory by eq. 3. }  
\label{kp}
\end{figure}

It is generally believed that the hadrons, after being formed during the confinement phase transition, interact on the way to the detector. Event generator are based on this assumption but the results of EPOS  \cite{Werner:2010aa} demonstrates that this rescattering changes the spectra only at low $p_t$, where almost no experimental data are available. EPOS succeeds to reproduce the measured spectra in between a factor of two and, in consequence, to reproduce the change of the spectral form from central to peripheral collisions.  The origin of this success, however, is little transparent due to the complexity of the approach.  Therefore it is worthwhile to study the spectra in an approach which contains the (for the success of EPOS) essential separation of core and corona particle but which is otherwise as simple as possible, i.e. in the core-corona model \cite{Aichelin:2008mi,Aichelin:2010ed,Aichelin:2010ns}. In this model we can calculate which spectra would be expected if no final state interactions among hadrons take place and we can use the difference to the data to learn something about the final state interaction. Assuming no final state interactions, in the core-corona model the spectra are superpositions of two contributions: the core contribution and the corona contribution.  The corona distribution  $ \frac{d^2N_i^{corona}}{2\pi . p_t . dp_t . dy}$ is the measured pp spectra divided by two, the core contribution  $ \frac{d^2N_i^{core}}{2\pi . p_t . dp_t . dy}$ is obtained from the experimental spectra for the most central AA collisions corrected for the corona contribution (see table I) and divided by the number of core participants. Then in the core-corona model the spectra for a given centrality is given by:
\be
\frac{d^2M_i}{2\pi . p_t . dp_t . dy}=
N_{part}[ (1 - f_{core}(N_{part}))\frac{d^2N_i^{corona}}{2\pi . p_t . dp_t . dy}+f_{core}(N_{part})\frac{d^2N_i^{core}}{2\pi . p_t . dp_t . dy}].
\ee
Fig. \ref{kp} presents the results for  $K^+$ mesons. We see in the top row the
spectra measured by the different experimental groups in comparison with the predictions of the core corona model.
The left hand side presents the STAR data, the middle panel the PHENIX data and  the right hand side the BRAHMS data. 
The middle row displays the difference between theoretical predictions and the data. The shaded region mark the error bars
(which are taken as the averaged error bar over the centrality bins).
We observe that for almost all STAR data points the theoretical predictions are in the experimental error bars. In peripheral reactions there is a tendency that at large $p_t$  the core-corona model is above the data. For semi central reactions model and data are in agreement for all $p_t$ values. This is not trivial at all as the bottom row shows. There we display the inverse slope parameter obtained by fitting the experimental and theoretical spectra by a thermal spectra 
\bea\frac{d^2N}{2\pi.m_t.dm_t.dy} = C.m_t.e^{\frac{-m_t}{T}}
\label{thermo}\eea
with $m_t$ being the \emph{transverse mass} of the considered particle:
\bea m_t = \sqrt{{p_t}^2 + m^2}\eea
where  $p_t$ is the \emph{transverse momentum} of the particle (with respect to the beam axis) and $m$ its free mass.
Even if the curves are not exactly exponential and therefore the value of the inverse slope parameters depends on fit range the bottom panel shows clearly that the slope varies considerably from central to peripheral reactions (which is in the core corona approach a consequence of the different invariant slope parameters in pp and central AA collisions). Also the PHENIX data are compatible with the core-corona model besides  the second last and third last centrality bin where the deviations, expected from  fig. 1, show up. We note in passing that for those centrality bins the core-corona models agrees well with
the STAR data. The central BRAHMS data are also compatible with the model but we see deviations for the most peripheral bin. It is remarkable that the peripheral STAR data are almost exponential whereas those of PHENIX are not and even less those of the BRAHMS collaboration. Comparing the three experimental spectra with the model we can conclude that for each experiment the majority of data is well described by the model. Deviations are specific for the experiment. There are no systematic deviations.

\begin{figure}
\includegraphics[width=0.3\textwidth]{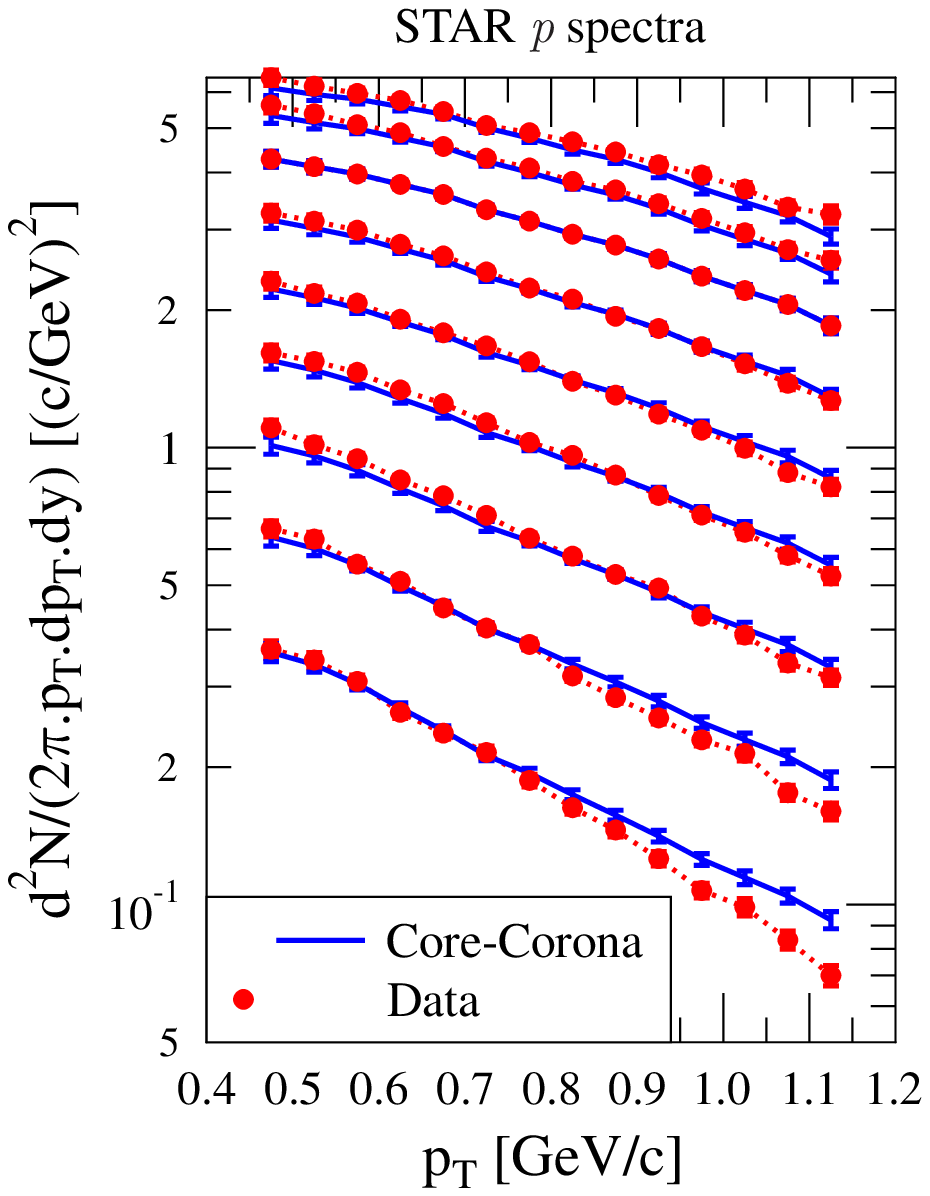}
\includegraphics[width=0.3\textwidth]{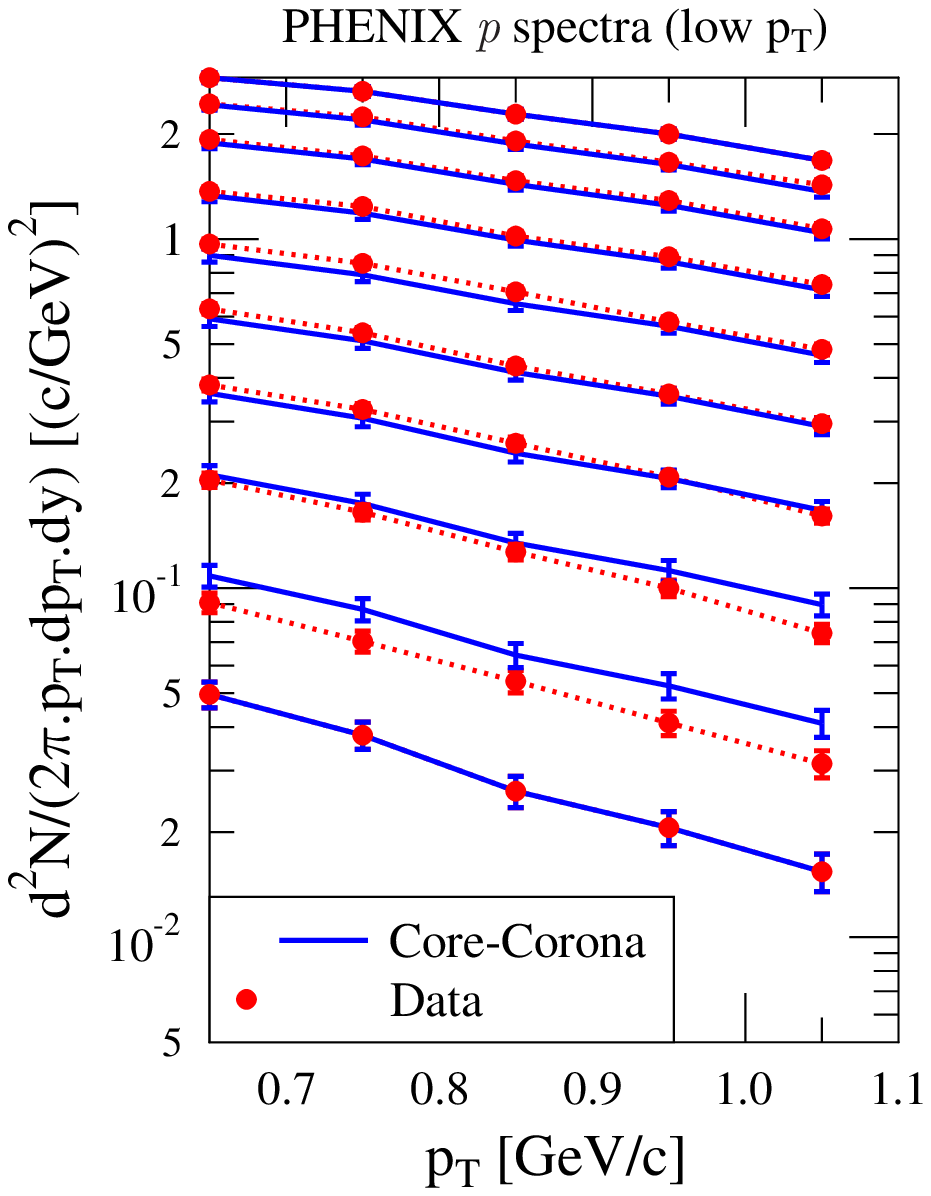}
\includegraphics[width=0.3\textwidth]{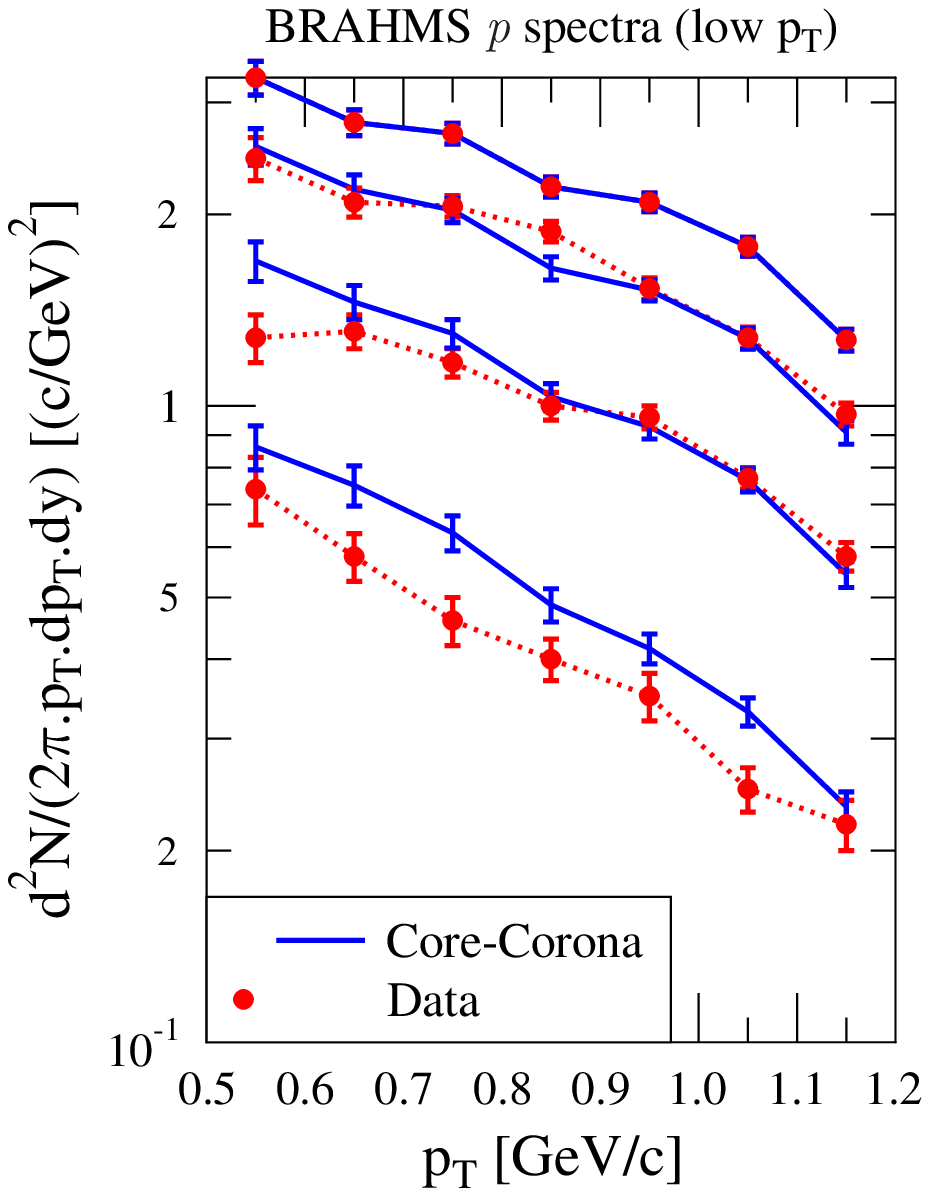}

\includegraphics[width=0.3\textwidth]{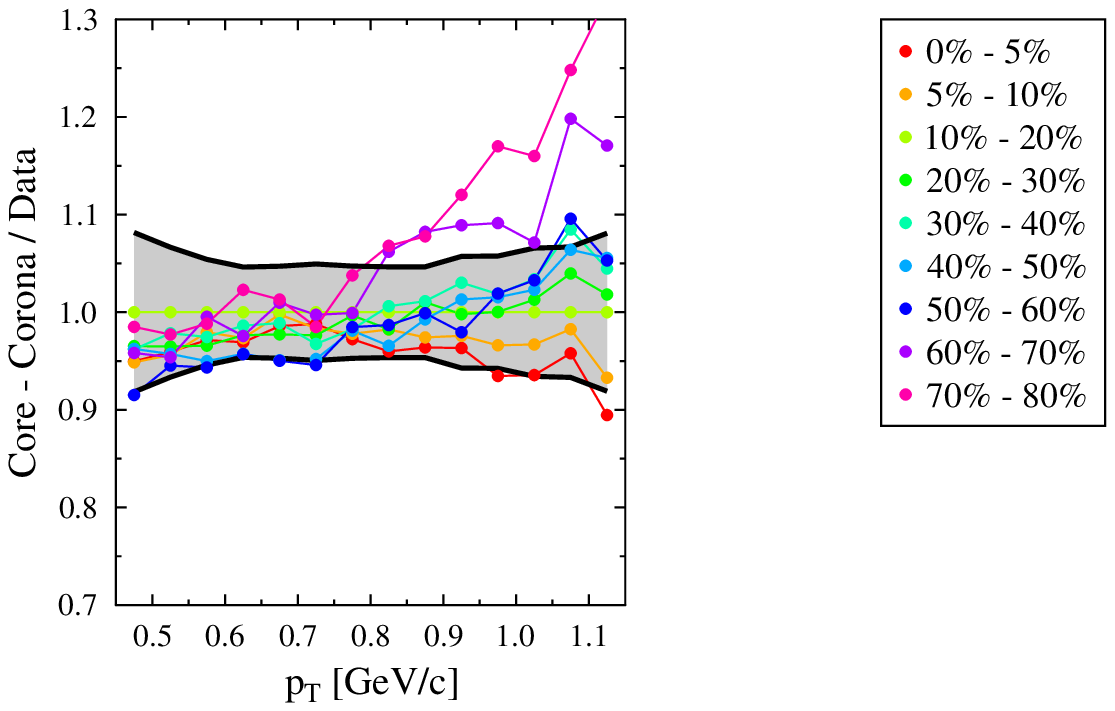}
\includegraphics[width=0.3\textwidth]{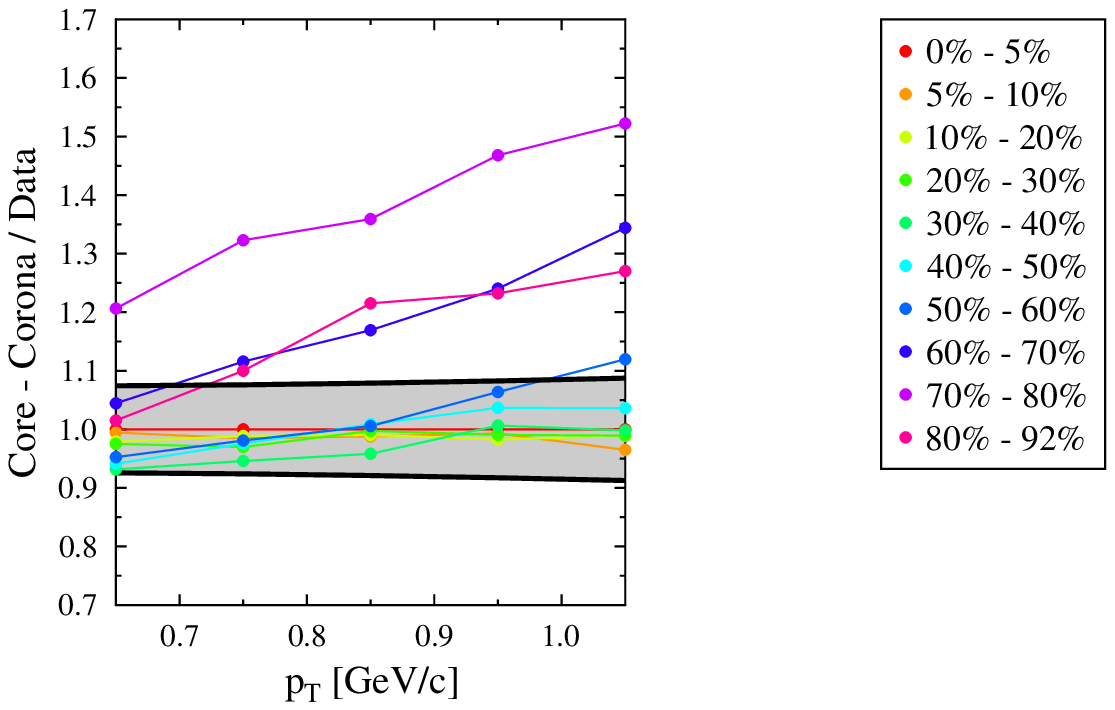}
\includegraphics[width=0.3\textwidth]{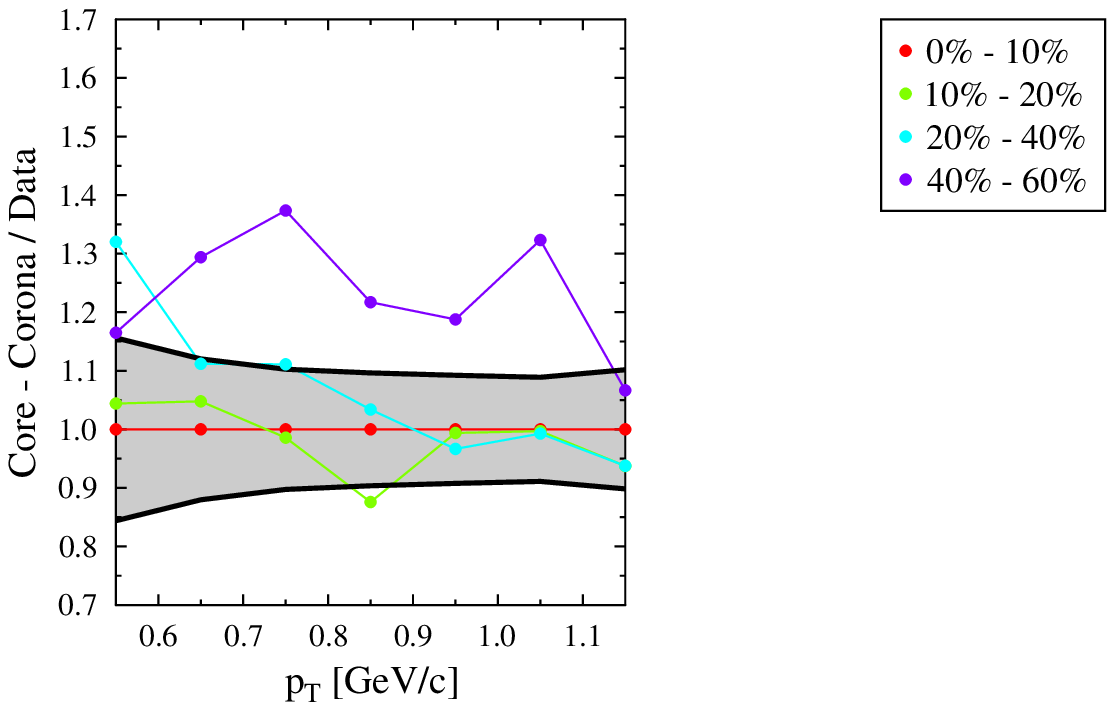}

\includegraphics[width=0.3\textwidth]{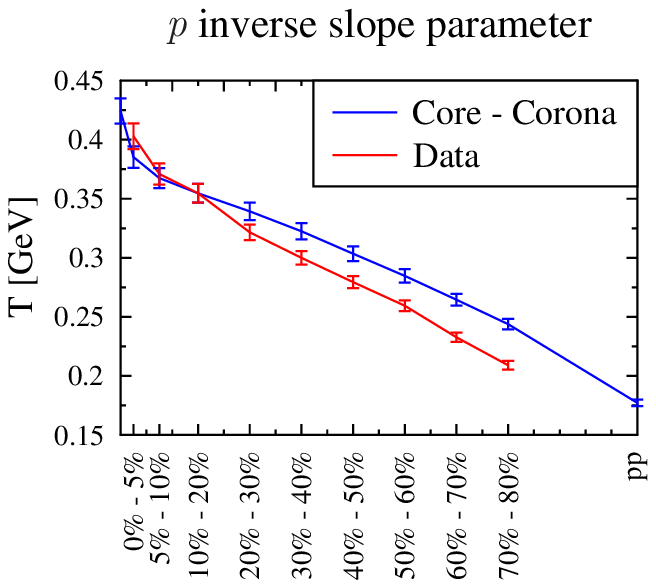}
\includegraphics[width=0.3\textwidth]{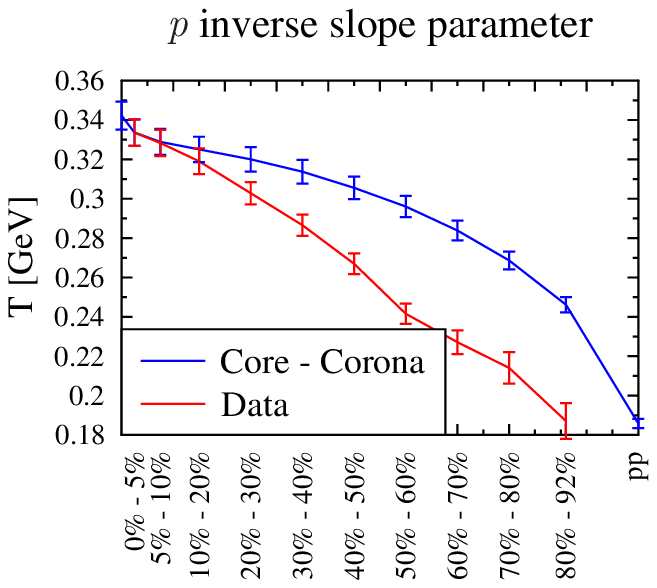}
\includegraphics[width=0.3\textwidth]{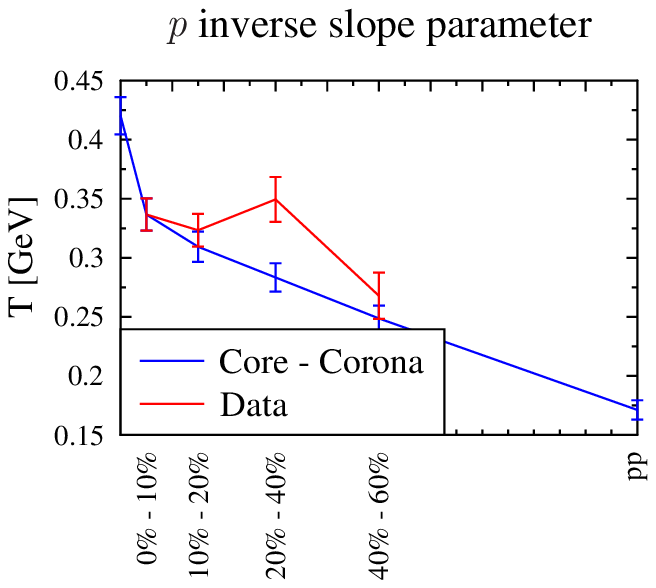}
\caption{Centrality dependence of the proton spectra measured by the STAR (left), PHENIX (middle) and BRAHMS (right) collaborations in comparison with the predictions of the core-corona model. Top:
The experimental spectra in comparison with the model prediciton. Middle: Ratio of the prediction of the
core-corona model and the experimental data. The shaded area mark the experimental errors. Bottom: Inverse slope parameter T obtained by fitting data and theory by eq. 3.}  
\label{pp}
\end{figure}

Fig. \ref{pp} show the same quantities for the protons.
For the STAR data see an almost perfect agreement between data and model predictions.
The only exceptions are, as for the $K^+$,  peripheral data at large $p_t$. The middle left figure demonstrates this in detail. We see that 
besides the high $p_t$ points in peripheral collisions the spectral form is reproduced by the core-corona model in between the error bars. This is far from being trivial. The inverse slope parameter of the spectrum varies by a factor of 2 between central
and peripheral reactions and so does the inverse slope parameter of the model due to the large difference between in the inverse slope parameters in pp and central AA. Almost the same is true for the PHENIX data. Here the model overpredicts the data of the second to last centrality bin by an almost constant factor. That this bin is particular we have already discussed in fig.1. In the BRAHMS proton data the form varies from central to peripheral reaction, what is not seen in the STAR and PHENIX data. This can also not be reproduced in the core-corona model. We do not display here the comparison for
the $\pi$, which is also reproduced in the core-corona model,  because neither the multiplicity per participant (see fig. 1) nor the slope varies considerable from central to peripheral reactions. Therefore it is not a good test ground for the validity of the core-corona model.  
\begin{figure}[h]
\centering
\includegraphics[scale=0.4]{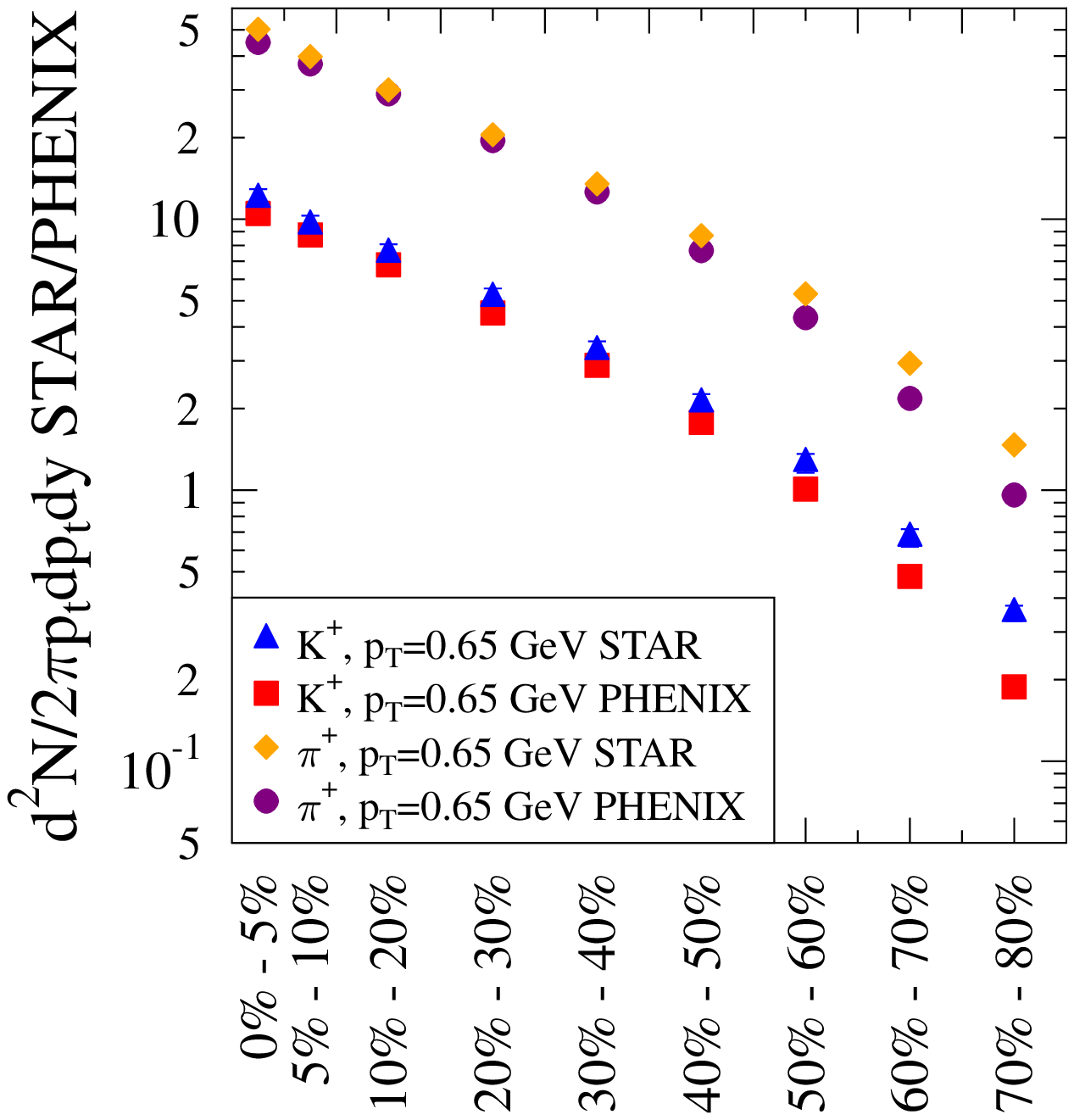}
\includegraphics[scale=0.4]{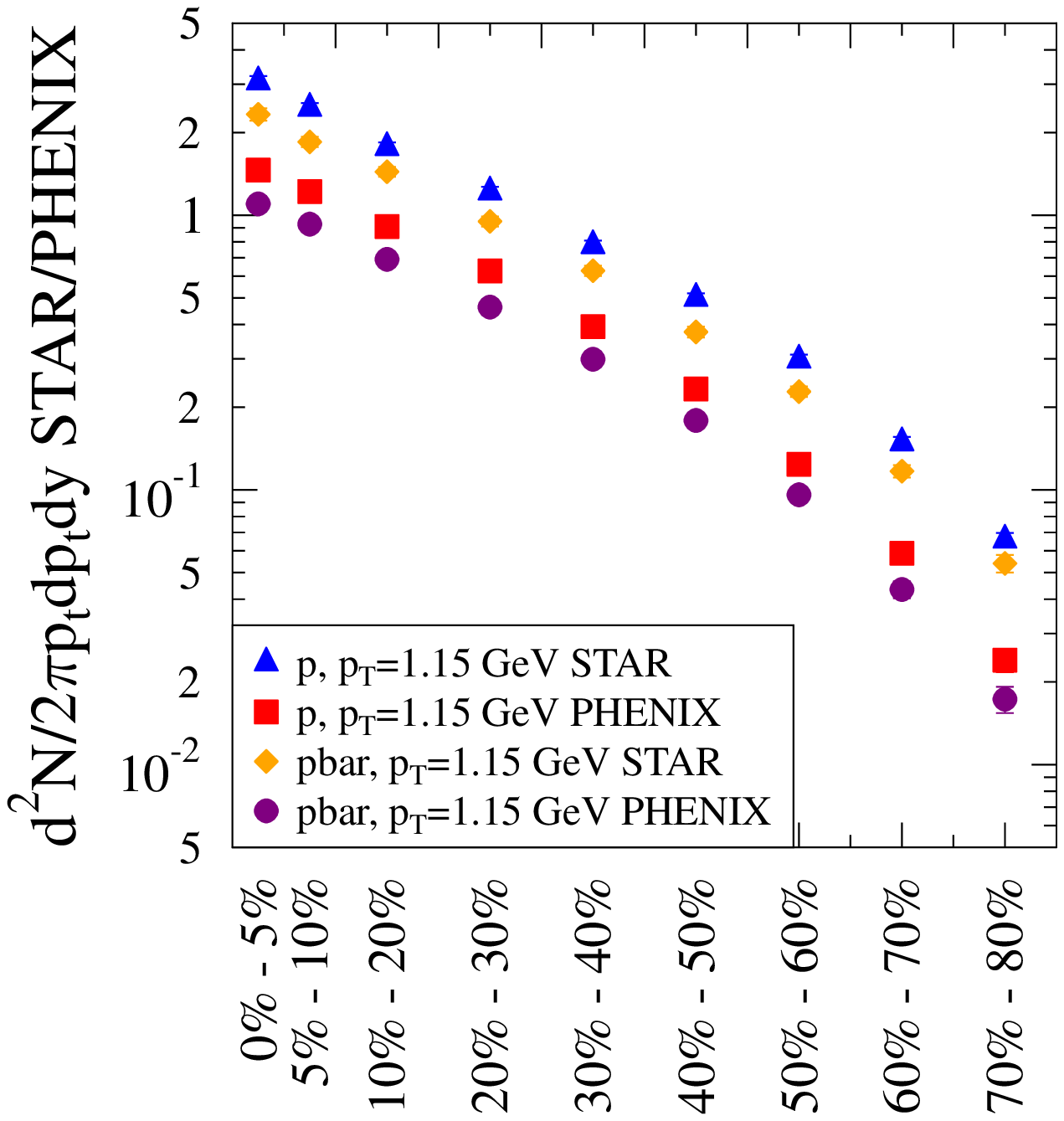}
\caption{Comparison of of the centrality dependence of the yield  measured by STAR and PHENIX 
for a given $p_t$.}
\label{spec2}
\end{figure}
\begin{figure}[h]
\centering
\includegraphics[scale=0.5]{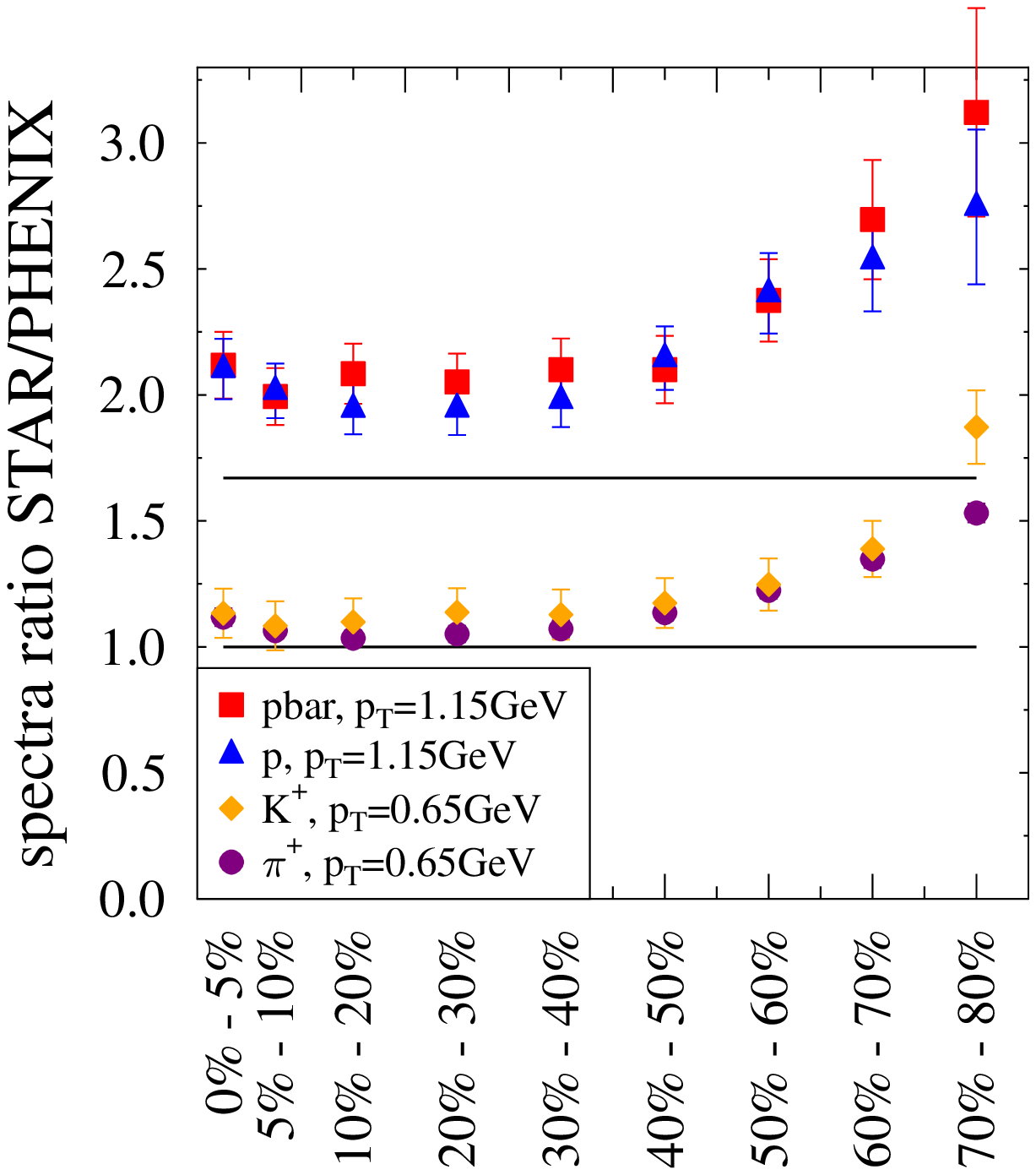}
\caption{Ratio of the yield for a given $p_t$  measured by STAR and PHENIX as a function of the centrality.}
\label{fig:pt}
\end{figure}

Thus the $K^+$ and $p$ spectra can be described in the core-corona model, however with rather different distributions  for the core and corona particles. Why different parameters are necessary is demonstrated in detail in figs.  \ref{spec2} and \ref{fig:pt} in which we compare directly the data of the PHENIX and of the STAR collaboration for a fixed value of $p_t$. 
Fig.\ref{spec2} shows how the yield of this $p_t$ bin develops as a function of the centrality. Fig. \ref{fig:pt} displays the ratio of the yields measured by STAR and PHENIX as a function of the centrality.

For the $\pi ^+$  and the $K^+$ the form of the spectra measured by STAR and by PHENIX deviate for peripheral reactions and for the $K^+$ in addition also for the most central collisions. Therefore the ratio of the spectra deviates from the expected value of 1 ( shown as a straight line in fig. \ref{fig:pt}). Only three out of the 18 points agree in the error bars
( which are added quadratically) with 1. 

The proton and antiproton spectra should not be identical because the PHENIX data have been corrected for weak decay whereas the STAR data are not. According to the PHENIX results this correction should be around
30  \% - 40\% . This would explain a ratio of 1.66 between the data shown as well as a straight line in fig \ref{fig:pt}. The data differ, however,  by more than a factor of two for central and semi central reaction and the difference increases toward peripheral reactions, as seen in fig. \ref{fig:pt}.  

In conclusion we have shown that all three available data sets for $p_t$ spectra of protons and kaons can be well described in the core-corona model. There are deviation but where they occur varies from experiment to experiment. Unfortunately no common parameter set can be found which describes the three experiments simultaneously. This is due to the strong differences between the experimental results for more peripheral events and due to the differences of the spectra measured in pp where the multiplicities differ up to  a factor of two between the three experiments.

The fact that the core-corona model describes also the single particle spectra of identified particles has important consequences:\\
a) The fact that the spectra at different centralities are only a superposition of that of core particles which are in thermal equilibrium and that of corona particles which have a spectrum as measured in pp collision testifies that there is little interaction between core and corona particles. Otherwise momentum would be transfered between the different particle species which have a quite different average $p_t$. This would, in consequence,  modify the spectral form.\\
b) The observed different centrality dependence of the average $p_t$ for different particles which has been interpreted as a signal for collective flow is merely due to the fact that the difference of $p_t$ in central AA and pp collisions is particle 
specific. The difference is large for protons,  practically inexistent for pions and that for kaons is in between. Of course
it may be that the collective flow is the origin of the different values for $p_t$ in central heavy ion and pp collisions. The core-corona model tell us that then the collective flow is independent of the size of the core part.\\
c) The blast wave fit, usually applied to characterize the data, is not the proper tool to analyze the spectra. Rather then describing a two component spectra by one single fit it would certainly be worthwhile to subtract the corona part in oder to study the properties of the core part which may carry the desired signal of the plasma.\\
Adding these observations to the already published analysis of the multiplicity \cite{Aichelin:2008mi}, $<p_t>$ and $v_2$ of identified particles \cite{Aichelin:2010ed, Aichelin:2010ns} in the core corona model one can conclude that presently all data on the centrality dependence of observables can be quantitatively described in this model. Predictions for the CuCu are straight forward and do not need any further input.
It remains to be seen whether other approaches based on non-viscous hydrodynamics yield a similar good description of the data. This would a prerequisite for validating the idea that the centrality depedence of $v_2/\epsilon$ is really due to a finite viscosity and not due to a distinction between core and corona particles.

Acknowledgment: We would like to thanks Dr. Roehrich and Dr. Yang for making us the Brahms pp data available and for an interesting discussion.

\end{document}